\documentclass[aps,pre]{revtex4-2}

\usepackage{graphicx}
\usepackage{amsmath}
\usepackage{amsfonts}
\usepackage{amssymb}
\usepackage{bm}
\usepackage{xcolor}
\usepackage{ulem}

\newcommand{\beq}{\begin{equation}}
\newcommand{\eeq}{\end{equation}}
\newcommand{\beqa}{\begin{eqnarray}}
\newcommand{\eeqa}{\end{eqnarray}}

\begin{document}

\title{Revisiting the Toda-Brumer-Duff criterion for order-chaos transition \\
in dynamical systems}

\author{F. Sattin}
\email{fabio.sattin@igi.cnr.it}
\affiliation{Consorzio RFX (CNR, ENEA, INFN, Universit\`a di Padova, 
Acciaierie Venete SpA), Corso Stati Uniti 4, 35127 Padova, Italy}
\author{L. Salasnich}
\email{luca.salasnich@unipd.it}
\affiliation{Dipartimento di Fisica e Astronomia ``Galileo Galilei'' 
and QTech, Universit\`a di Padova, via Marzolo 8, 35131 Padova, Italy}
\affiliation{Istituto Nazionale di Fisica Nucleare, Sezione di Padova, 
via Marzolo 8, 35131 Padova, Italy}
\affiliation{Istituto Nazionale di Ottica del Consiglio Nazionale delle 
Ricerche, Unit\`a di Sesto Fiorentino, via Carrara 1, 50019 
Sesto Fiorentino (Firenze), Italy}

\begin{abstract}
The Toda-Brumer-Duff (TBD) is an analytical criterion for estimating the local exponential rate of divergence between nearby trajectories in dynamical systems, and it is employed as a test for assessing the existence of chaos therein. 
It is fairly simple, intuitive, and works well in several situations, 
hence gained quite a wide popularity, yet it is known to be not rigorous 
since predicts ``false positives'', i.e., flags as chaotic systems that are 
instead regular. We revisit here the TBD criterion in order to understand the causes of its failures, 
and pinpoint that the problem with it is due to two reasons: (a) the TBD criterion does not 
constrain the trajectories to lie on the same energy hypersurface; 
(b) it does not 
distinguish between the divergence of trajectories along or perpendicularly to the direction 
of the flow, the former being irrelevant for assessing the presence of chaos. 
We show how both points can be incorporated within the TBD framework, 
yielding an amended criterion which, when 
applied to some reference cases, interprets correctly the kind of dynamics 
observed.

\hfill


\end{abstract}

\maketitle

\section{Introduction}
Dynamical systems may feature both regular or irregular chaotic motion depending on control parameters. Quite often one such parameter is the energy $E$: at low $E$ the system is usually regular, and a smooth or sharp transition to irregular motion occurs by increasing $E$. Identifying the kind of behavior of a dynamical system is obviously extremely important; it can be done directly through the estimate of the Lyapunov exponents or the Poincar\'e plots, which are the two most known and employed numerical techniques (see, for instance, Refs. \cite{book0-chaos,book1-chaos}). In the course of time, several other algorithms have been developed \cite{book2-chaos}, including the Smaller Alignment Index (SALI) \cite{sali}, the analysis of the power spectrum of geodesic divergence between nearby trajectories \cite{geodes}, the Fast Lyapunov Indicator method \cite{fli}, 
and methods applying directly to time series \cite{pd2005}. 
More recent proposals involve the Shannon entropy \cite{pd2021} and non-variational Lagrangian methods based on Lagrangian
Descriptors \cite{pd2022,pd2023}.  All these techniques require the numerical integration of the equations of motion for several trajectories. In the past, when computer resources were more scarce, there was a premium for overriding this computationally expensive step and assessing the regular/irregular behavior through the analytical inspection of the structure of the equations of motion. Even nowadays, analytical approaches may be valuable, since provide insight into the mechanisms that generate chaos and the paths that drive the order/chaos transition. One such algorithm, valid for two-dimensional systems not too far from integrable ones, is provided by Codaccioni, Doveil and Escande \cite{codaccioni82}. 

The first analytical approach was developed in 1970 by Okubo \cite{okubo70}, within the framework of fluid mechanics, but it is nowadays more widely known from the works of Toda \cite{toda74}, Brumer and Duff \cite{brumer76}, applied to generic Hamiltonian dynamical systems.  Within this work, we will be restricted to Hamiltonians of the most common kind, with separated kinetic and potential terms (and with unitary mass for simplicity)
\begin{equation}
H =  { |{\bf p}|^2 \over 2 } + V({\bf q}) \; , 
\label{eq:1}
\end{equation}
where the potential $V({\bf q})$ is a nonlinear function of the $N$ 
coordinates ${\bf q}=(q_1,q_2,...,q_N)$ and ${\bf p}=(p_1,...,p_N)$ 
is the vector of linear momenta. 
Hamilton equations  write
\beqa
{\dot {\bf q}}(t) &=& {\bf p}(t) \; , 
\label{Eq:2a}
\\
{\dot {\bf p}}(t) &=& - \partial_{{\bf q}} V({\bf q}(t)) \; , 
\label{Eq:2b}
\eeqa
where the dot means the derivative with respect to time $t$ 
and $\partial_{\bf q}$ means the gradient with respect to ${\bf q}$. \\
The stability of the motion is investigated by looking at two nearby trajectories
$({\bf q}_1, {\bf p}_1)$ and $({\bf q}_2, {\bf p}_2)$ and rewriting the Hamilton 
equations in terms of the displacements 
$\delta {\bf q} = {\bf q}_1 - {\bf q}_2$, $\delta {\bf p} = {\bf p}_1 - {\bf p}_2$ as follows
\begin{eqnarray}
\delta {\dot {\bf q}}(t)  &=& \delta {\bf p}(t)  
\label{Eq:3a} 		\\
\delta {\dot {\bf p}}(t)  &=& -  \left[ \partial_{\bf q} V(\delta {\bf q}(t) 
+ {\bf q}_2(t)) - \partial_{\bf q} V({\bf q}_2(t)) \right] \; . 
\label{Eq:3b}
\end{eqnarray} 
Defining the $2N$-dimensional 
displacement vector ${\bf D}=(\delta {\bf q},\delta{\bf p})$, 
the right-hand sides of the previous equations are linearized 
around ${\bf D} = 0$: 
\beq
{\dot {\bf D}}(t) = {\hat{\bf M}(t)} \ {\bf D}(t) \; , 
\label{Eq:4}
\eeq
where 
\begin{eqnarray} 
\hat{\bf M}(t) &=&   
\begin{pmatrix} 0 & \hat{\bf I}_N \\
- \hat{\bf H}& 0 
\end{pmatrix}  \\
(\hat{\bf I}_N)_{ij} &=& \delta_{ij} \; , \quad (\hat{\bf H})_{ij} = \partial^2_{ij}  V({\bf q}(t))  \nonumber
\end{eqnarray}
Notice that $\hat{\bf M}$ depends from time only implicitly, through the coordinates ${\bf q}$'s which evolve by virtue of Eqns. (\ref{Eq:2a}, \ref{Eq:2b}). 
The solution of the coupled system (\ref{Eq:2a},\ref{Eq:2b},\ref{Eq:4}) over long times yields the largest Lyapunov exponent of the system, a positive value flagging the existence of chaos. As it is well known, the estimate of Lyapunov exponents in most cases can only be done numerically.
Toda, Brumer and Duff (TBD) postulate that a scale separation holds: 
the coordinates of the common motion $({\bf q}, {\bf p})$ 
are considered as frozen while evolving ${\bf D}(t)$. That is, 
the entries of the $2N \times 2N$ matrix $\hat{\bf M}$ are considered 
as constant while solving Eq. (\ref{Eq:4}), which decouples from Eqns (\ref{Eq:2a},\ref{Eq:2b}) and has exponential solutions
\begin{equation}
{\bf D}(t) = \sum_{i=1}^{2 N} \chi_i {\bf d}_i \exp(\mu_i t)
\label{Eq:mu}
\end{equation}
where $\mu_i$ are the eigenvalues of the matrix $\hat{\bf M}$, ${\bf d}_i$ the 
corresponding eigenvectors and $\chi_i$ numerical coefficients depending upon initial conditions. 
Now the coordinates ${\bf q}$'s appearing in the entries of the matrix $\hat{\bf M}$ (and therefore also in $\mu_i$)
play the role of arbitrary parameters, have no necessary reference to the particle's trajectory. 
Positive real-valued eigenvalues $\mu_i$ 
imply local exponential divergence between neighboring trajectories at the position $\bf q$. 
TBD argue that the regions where exponential 
divergence exists are the seeds for the global chaotic behavior of the 
system. The rationale is that orbits visiting repeatedly regions where at least one $\mu_i > 0$ become unstable and ultimately chaotic. 

The TBD criterion is extremely intuitive and simple to implement; furthermore, it proved effective in predicting order-chaos transitions for several systems, thus it became quite popular. Besides generic Hamiltonian systems \cite{enz75,nunez90,oloumi99}, it was applied to problems from field theory \cite{watabe95, kuvnishov02,kuvnishov08}, atomic physics \cite{li95}, nuclear physics \cite{manfredi97, letelier11}, plasma physics \cite{ghosh14}, neutral fluid and MHD dynamics \cite{chang14,beron19,shivamoggi22}.
It was soon realized, though, that it is not a rigorous criterion. Benettin et al. \cite{benettin77} showed that it does not exist a clear correspondence between the loci of instability predicted by the TBD criterion and the actual exponential departure of numerically computed trajectories. Furthermore,  the TBD criterion predicts ``false positives'', i.e., regions of hyperbolic instability in systems that can be rigorously proved to be regular, including one-dimensional systems like the pendulum, or two-dimensional systems like the anti-Hènon-Heiles system \cite{benettin77,pattanayak97,smirnov98}. These difficulties are way more remarkable given the streamlined derivation of the TBD criterion. 
It is obvious that the problem arises when decoupling Eq. (\ref{Eq:4}) from Eqns. (\ref{Eq:2a},\ref{Eq:2b}), but it is not easy to identify where it lies.
The purpose of this paper is (i) identify the pitfalls of the TBD criterion and (ii) establish whether it is possible to get rid of them, safeguarding at the same time its mathematical simplicity, at least to some extent. The next section provides a re-examination of the TBD criterion, answering the question (i). Section 3 presents an implementation of the amended criterion to low-dimensional smooth dynamical systems, in order to address point (ii), and Section 4 reports its performances when applied to a few selected reference models, which are popular standards in literature.

\section{Scrutiny of the TBD criterion}

The realization that the TBD criterion was not exact led immediately to attempts to improve it. The first proposal was put forth by Cerjan and Reinhardt \cite{cerjan79}. They argued that the linearization step leading to
Eq. (\ref{Eq:4}) actually entails investigating the stability properties of the fixed point $\dot{ \bf D} = 0 $ at ${\bf D} = 0$. If, besides this one, further stable fixed points at finite values  $\bf D$ do exist, any initially unstable perturbation will be ultimately dragged into the attraction basin of one of these points as it grows. The Cerjan-Reinhardt suggestion is clever but, basically, is a model for the nonlinear saturation of an instability. It does not explain why the initial stages of the instability do not produce observable effects, in the first place. 

In 1981, Kosloff and Rice \cite{kosloff81} realized that the displacement vector ${\bf D}$ within the TBD criterion is totally arbitrary, whereas two conditions are to be fulfilled: 
(i) Both trajectories $({\bf q}_{1}(t), {\bf p}_{1}(t))$ and 
$({\bf q}_{2}(t), {\bf p}_{2}(t))$ lie on the constant 
energy hypersurface $H = E$, thus ${\bf D}(t)$ must be tangent to the 
same hypersurface. 
(ii) The vector ${\bf D}(t)$ may be decomposed as 
\beq 
{\bf D}(t)= {\bf D}_{\parallel}(t) + {\bf D}_{\perp}(t) \; , 
\eeq
where ${\bf D}_{\parallel}(t)$ is parallel to the flow: ${\bf D}_{\parallel}(t) \parallel (\dot{\bf q}, \dot{\bf p})$, and ${\bf D}_{\perp}(t)$ is 
perpendicular to it (still lying on the surface $H=E$). Only the exponential growth of ${\bf D}_{\perp}(t)$ is relevant in establishing 
the chaotic properties of trajectories: choosing two initial points 
along the flow is equivalent to considering one and the same trajectory, 
just starting at different times. Within the TBD criterion, there is no way 
for discriminating whether exponential growth takes place along ${\bf D}_{\parallel}(t)$ or ${\bf D}_{\perp}(t)$.
 
It is straightforward to realize that accounting for the two conditions 
(i) and (ii) rule out all one-dimensional systems ($N=1$) from the 
consideration of the TBD criterion, since it is not possible 
to write couples of isoenergy trajectories having at the same time 
${\bf D}_{\perp}(t) \neq {\bf 0} $. 

In the following, we will consider the specific but important case of two-dimensional systems ($N=2$). We will show 
how a correct consideration of the two conditions (i, ii) leads to a modified TBD criterion. 
There are no conceptual obstacles preventing the same analysis from being applied to higher-dimensional systems, but the computational burden grows accordingly, as well as the difficulty of visualizing and interpreting the results.

\subsection{Two-dimensional systems:  $N = 2$}

From now on we will use the symbols $(x,y)$ rather than $q_1, q_2$, in 
adherence with customary usage and in order not to burden notations. 

Let us define the following local orthonormal basis system in the phase-space: 
\begin{eqnarray}
{\bf e}_0 &=& \left( \partial_x V, \partial_y V, p_x, p_y \right)  Q^{-1} \\
{\bf e}_1 &=&  \left(p_x,  p_y, -\partial_x V, -\partial_y V \right)  Q^{-1}\\
{\bf e}_2 &=& \left(-p_y, p_x, -\partial_y V, \partial_x V \right)  Q^{-1}\\
{\bf e}_3 &=& \left(\partial_y V, - \partial_x V, -p_y, p_x \right)  Q^{-1} 
\label{Eq:5}
\end{eqnarray} 
where $Q =  \sqrt{(\partial_x V)^2 + (\partial_y V )^2 + {\cal{K}}^2}$, and $ {\cal{K}} = \sqrt{p_x^2+p_y^2}= \sqrt{2 (E - V)}$. 
Energy surfaces are defined by the condition
\begin{equation}
E = H = {p_x^2 \over 2} + {p_y^2 \over 2} + V(x,y) 
\label{Eq:5.1}
\end{equation}
The direction perpendicular to the energy surface is defined by the phase-space gradient of $H: (\partial_q H, \partial_p H) $; it is straightforward to check that this direction is spanned by ${\bf e}_0$. 
The second vector is aligned along the flow: ${\bf e}_1 \propto \left( {\dot x}, {\dot y}, {\dot p}_x , {\dot p}_y \right) = \left(p_x,  p_y, -\partial_x V, -\partial_y V \right) $. Finally, the two vectors ${\bf e}_2, {\bf e}_3$ span the remaining two directions perpendicular to the flow.   

We write the displacement vector as a linear combination over the basis (\ref{Eq:5}), restricting to the hyperplane tangent to the energy surface: that is, no component along ${\bf e}_0$ is allowed. Thus, we write ${\bf D}$ in Eq. (\ref{Eq:4}) to be of the form
\begin{equation}
{\bf D}(t) = \sum_{i=1}^3  \alpha_i(t) {\bf e}_i . 
\label{Eq:5.11}
\end{equation}

Consistently with the logic of the TBD approach, we keep the basis ${\bf e}_i$ as fixed and let the coefficients $\alpha_i$ alone vary in time.  We replace this expression into Eq. (\ref{Eq:4}) which project onto the basis vectors ${\bf e}_1,  {\bf e}_2, {\bf e}_3$.
We introduce the array  $\bm{\alpha} = (\alpha_1,   \alpha_2, \alpha_3)$ and 
the $3 \times 3$ matrix $\hat{\bf W}$:
$ {W}_{ij} =  {\bf e}_i \cdot \hat{\bf M} \cdot {\bf e}_j$. 
Thus, we started from the $4 \times 4 $ array equation (\ref{Eq:4}) and 
arrived to the $ 3 \times 3$ one 
\begin{equation}
{\dot {\bm{\alpha}}}(t) = \hat{\bf W} \ {\bm{\alpha}}(t)
\label{Eq:5.2a}
\end{equation}  
We reiterate that the entries of the matrix $\hat{\bf W}$ depend on the phase-space coordinates $(x,y,p_x,p_y)$ which must here be considered as parameters.    
The solution of (\ref{Eq:5.2a}) with given initial 
condition $\bm{\alpha}(t=0) = \bm{\alpha}_0$ is
\begin{equation}
\bm{\alpha}(t) = \sum_{i=1}^3 (\bm{\alpha}_0\cdot{\bf w}^{(i)}) \, 
\exp(\lambda^{(i)} \, t) \, {\bf w}^{(i)} 
\label{Eq:8}
\end{equation}
The  $\lambda^{(i)}$'s are the eigenvalues of $\hat{\bf W}$, roots of 
the third-order polynomial
\begin{equation}
det(\hat{\bf W} - \lambda \hat{\bf I} ) =    -\lambda^3 + c_2 \lambda^2 + c_1 \lambda + c_0 = 0 
\label{Eq:7}
\end{equation}
with
\begin{eqnarray}
c_2&=& { p_x \partial_x V (\partial_{xx} V -1)  +  p_y \partial_y V (\partial_{yy} V -1) + \partial_{xy} V (p_y \partial_x V + p_x \partial_y V) \over Q^2 } , \nonumber \\
c_1&=& - {\partial_{xx}V (p_y^2 + \partial_y V^2)  + \partial_{yy}V (p_x^2 + \partial_x V^2) - 2 \partial_{xy} V (p_x p_y + \partial_x V \partial_y V) \over Q^3 },  \nonumber \\
c_0&=& - { p_x \partial_x V ( \partial_{xy} V^2 + \partial_{yy} V - \partial_{xx} V \partial_{yy} V) +  
                 p_y \partial_y V ( \partial_{xy} V^2 + \partial_{xx} V - \partial_{xx} V \partial_{yy} V)  -
                 \partial_{xy} V ( p_y \partial_x V + p_x \partial_y V) \over Q^4 }\nonumber 
\label{Eq:7a}                 
\end{eqnarray}
where $Q$ was defined in correspondence of Eqns. (\ref{Eq:5}). Finally, the ${\bf w}^{(i)}$'s are the eigenvectors associated to the eigenvalues $\lambda^{(i)}$'s. 
The eigenvalues are the roots of the third-order polynomial with real-valued coefficients (\ref{Eq:7}), thus we know that one of the $\lambda$'s  must be purely real (but it may be either negative or positive). We will call it $\Lambda_R$. The other two eigenvalues are complex-valued, conjugated to each other: $\Lambda_I = \Lambda_{I,r}+ i \Lambda_{I,i}, \Lambda_I^* = \Lambda_{I,r}- i \Lambda_{I,i}$. The three corresponding eigenvectors will be labeled as ${\bf w}_R, {\bf w}_I, {\bf w}_I^*$. In conclusion, Eq. (\ref{Eq:8}) becomes
\begin{eqnarray}
\bm{\alpha}(t) &=& (\bm{\alpha}_0\cdot{\bf w}_R) \exp(\Lambda_R \, t) {\bf w}_R + 
2 Re\left[ (\bm{\alpha}_0\cdot{\bf w}_I) \exp(\Lambda_I \, t) {\bf w}_I \right] \nonumber \\
 &=&  (\bm{\alpha}_0\cdot{\bf w}_R) \exp(\Lambda_R t) {\bf w}_R + 
2  \exp(\Lambda_{I,r}\, t) \, \left\{\cos(\Lambda_{I,i} \, t) Re  \left[ (\bm{\alpha}_0\cdot{\bf w}_I)  {\bf w}_I \right] - \sin(\Lambda_{I,i} \, t) Im  \left[ (\bm{\alpha}_0\cdot{\bf w}_I)  {\bf w}_I \right] \right\} \nonumber \\
\label{Eq:8a}
\end{eqnarray}
At this point, the reader may get confused, so it is worthwhile a brief recapitulation of the different entities introduced so far:
\begin{itemize}
\item  $\bf D$, ${\bf e}_i$ are vectors in the physical phase-space; in particular, they have physical dimensions. 
\item Conversely, $\bm \alpha$, ${\bf w}_{R,I}$ are elements of an abstract space: they are pure numbers. 
\end{itemize}
In order to make further progress, we recall the second point raised by Kosloff and Rice: The existence of real-valued positive eigenvalues is not a sufficient condition to infer instability. 
We can envisage two potential instabilities corresponding to either $\Lambda_R   > 0 $ or $Re(\Lambda_I)  > 0$. 
To fix ideas, we first focus on the first case. The term proportional to $\exp (\Lambda_R \, t)$, therefore, dominates over the others in Eq. (\ref{Eq:8}) (but for particular choices of the initial conditions), thus the whole sum  ultimately must collapse onto just this single term 
\begin{equation}
\bm{\alpha}(t) \simeq  (\bm{\alpha}_0\cdot{\bf w}_R) \exp(\Lambda_R \, t) \, {\bf w}_R 
\label{eq:d1}
\end{equation}
We insert this expression in Eq. (\ref{Eq:5.11}):
\begin{equation}
{\bf D}(t) = \sum_{i=1}^3  \alpha_i(t) {\bf e}_i  \simeq \sum_{i=1}^3  (\bm{\alpha}_0\cdot{\bf w}_R) \exp(\Lambda_R \, t) \, ({\bf w}_R)_i \, {\bf e}_i
\label{eq:d2}
\end{equation}
The dilatation along the flow, $D_\parallel$, is given by the projection along ${\bf e}_1$: 
\begin{equation}
D_{\parallel}(t) = {\bf D}(t)\cdot {\bf e}_1 \simeq (\bm{\alpha}_0\cdot{\bf w}_R) \exp(\Lambda_R \, t) ({\bf w}_R)_1  
\end{equation}
Since ${\bf w}_R$ is a unit vector, the dilatation perpendicular to the flow is
\begin{eqnarray}
D_{\perp}(t)  &=& \sqrt{ ({\bf D}(t)\cdot {\bf e}_2)^2 +  ({\bf D}(t)\cdot {\bf e}_3)^2}  \nonumber \\
 &\simeq&  |\bm{\alpha}_0\cdot{\bf w}_R| \exp(\Lambda_R \, t) \sqrt{({\bf w}_R)_2^2+({\bf w}_R)_3^2} \nonumber \\
 &\equiv&  |\bm{\alpha}_0\cdot{\bf w}_R| \exp(\Lambda_R \, t) \sqrt{1-({\bf w}_R)_1^2}
\label{eq:d3}
\end{eqnarray}
Let us now investigate the second possibility: $Re(\Lambda_I)  = \Lambda_{I,r} > 0$. In this case, dilatation is modulated by the terms $\cos ,  \sin( \Lambda_{I,i} t) $, see Eq. (\ref{Eq:8a}). If $|\Lambda_{I,i} | \gg \Lambda_{I,r}$ the effective dilatation averages to zero over times of order $\Lambda_{I,r}^{-1}$, and thus in this case chaotic motion cannot arise as well. A demonstration that the condition  $|\Lambda_{I,i} | \gg \Lambda_{I,r}$ holds when $ \Lambda_{I,r} > 0 $ does not exist since it is not generically true, but we will numerically prove below that it is the case for all the test cases studied.  \\
We therefore conjecture the following test for chaotic motion: (i) one has first of all to identify potentially unstable regions where exponential dilatation of neighboring trajectories occurs. (ii) Secondly, within these regions, one estimates the associated eigenvector as a proxy for the fraction of dilatation that takes place along a direction parallel to the local flow. This dilatation does not contribute to chaotic motion therefore, if it is dominant--i.e. $|({\bf w}_R)_1|$ is fairly close to unity--we may argue that dynamics is regular even in the presence of positive eigenvalues.  
In summary: chaotic motion occurs provided that the following two conditions are simultaneously fulfilled over a fraction of the available phase-space: $\Lambda_{R} > 0$ {\it and}  $|({\bf w}_R)_1|$ appreciably smaller than unity. 
We are going to test this conjecture over some test models in the next section.

\section{Analysis of some two-dimensional test cases}

We consider here three study cases: 
(A) The Hènon-Heiles potential, and (B) The anti-Hènon-Heiles potential.
We choose them since they are classical reference cases for chaotic and regular motion. (C) To them, we added a third chaotic potential designed by ours, in order to enlarge the database. 

The investigative approach follows the logic of the TBD one. For each fixed energy$E$, we sample a large number of points $(x,y,p_x,p_y)$ in the allowed phase space, compute the corresponding eigenvalues $\Lambda$ and eigenvectors {\bf w}, and check if they are consistent with the criteria developed in the previous section for distinguishing between chaotic and ordered motion.

\subsection{The Hènon-Heiles system}

The Hènon-Heiles potential has the form
\begin{equation}
V = {1 \over 2} (x^2 + y^2) + x^2 y - {1 \over 3} y^3
\label{Eq:9}
\end{equation}
This system is known to feature an order-chaos transition at energy $ E \approx 0.083 $, and to be totally chaotic when $E = 1/6 \simeq 0.1667$ (above this value the equipotential curves are no longer closed. We will consider here only bounded systems) \cite{toda74,pattanayak97,zotos15}.  

\begin{figure}[h]
\includegraphics[width=10.5cm]{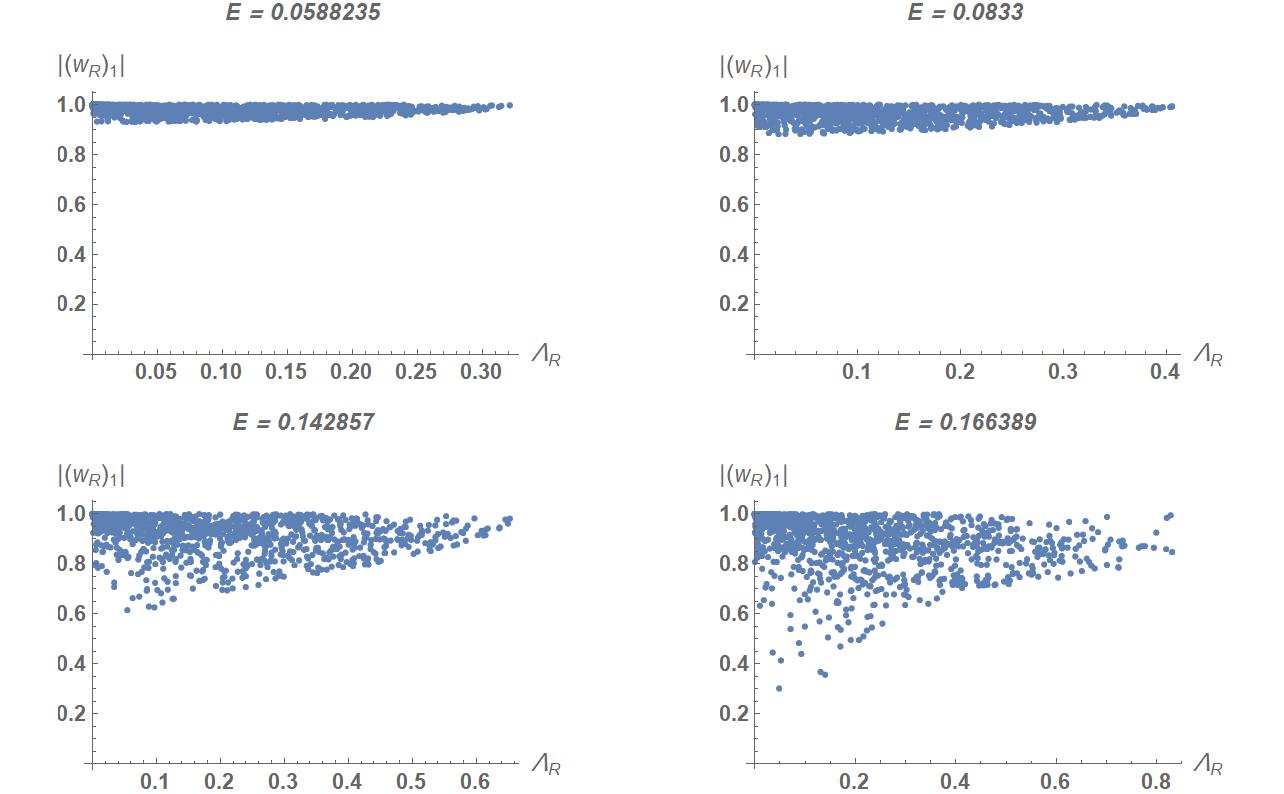}
\caption{$|({\bf w}_R)_1| $ {\it versus } $\Lambda_R$ plots, at different energies for the Hènon-Heiles system. For each plot,  a  total of 3000 points were randomly sampled within the allowed phase space. The  two plots  with lower energy feature a cloud of points which aligns fairly close to the horizontal straight line $|({\bf w}_R)_1| = 1$, consistent with our hypothesis concerning regular motion. Conversely, the two plots at the higher energies, in the chaotic domain, feature an appreciable scatter of the points throughout the $|({\bf w}_R)_1| < 1$ region.  }
\label{fig:1}
\end{figure}

Figure \ref{fig:1} reports the results corresponding to four values of the energy. For each plot, the horizontal axis contains the positive real-valued eigenvalue $\Lambda_R$, the vertical axis the projection of the corresponding {\bf w} eigenvector over the parallel direction, $({\bf w}_R)_1$. The lowest energy case is deep in the regular region,the next case is chosen exactly at the threshold for the onset of chaos, and the remaining two examples are taken in the chaotic window.
In correspondence of the smallest energy, all points align fairly close to the $|({\bf w}_R)_1| = $ line, with just a small spread, consistent with our picture of regular motion. A the onset of chaos, at $E = 0.083$, things are not qualitatively changed, obe observes just a mild spreading of the cloud of points, which remain however located above $\simeq 0.9$.  
When one enters the chaotic region, though, the spreading towards  smaller $|({\bf w}_R)_1|$ becomes more and more noticeable. 

\begin{figure}[h]
\includegraphics[width=10.5cm]{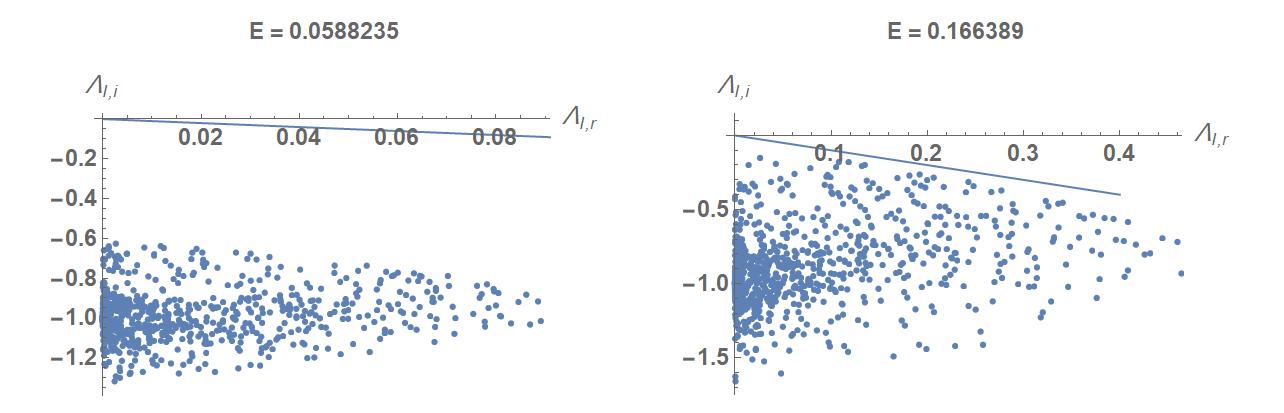}
\caption{The horizontal axis reports the real part of the complex-valued eigenvalue, $\Lambda_{I,r}$, when it is positive. The vertical axis reports the corresponding imaginary part, $\Lambda_{I,i}$. The straight line is $\Lambda_{I,i} = - \Lambda_{I,r}$. The data refer to the cases with the smallest and the largest energy out of those considered in Fig. \ref{fig:1}.  All the points lie in the region $|\Lambda_{I,i}| > \Lambda_{I,r}$, supporting our view that the effective stretching of these trajectories is negligible.  }
\label{fig:2}
\end{figure}

In Fig.\ref{fig:2} we present, limited to the smallest and largest energy, the plots $\Lambda_{I,i}$ {\it versus} $\Lambda_{I,r}$. 
In both cases, $|\Lambda_{I,i}| > \Lambda_{I,r}$, supporting our claim that the effects of the complex-valued eigenvalues upon the stretching of the trajectories is negligible.

\subsection{The anti-Hènon-Heiles system}
\begin{figure}[h]
\includegraphics[width=8.5cm]{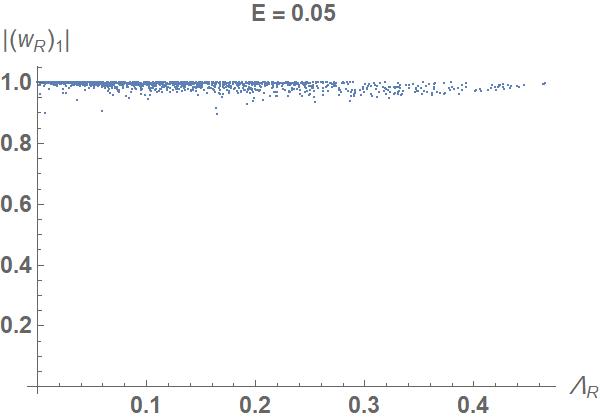}
\caption{The figure is the counterpart of Fig. \ref{fig:1} for the anti-Hènon-Heiles system. The energy is $E = 0.05$, i.e, above the threshold for the onset of chaos according to the TBD criterion. As expected on the basis of our hypothesis, all points lie fairly close to the $|({\bf w}_R)_1|  = 1$ curve. }
\label{fig:3}
\end{figure}

The potential differs from the previous case just by the sign of the cubic term in $y$:
\begin{equation}
V = {1 \over 2} (x^2 + y^2) + x^2 y + {1 \over 3} y^3
\label{Eq:10}
\end{equation}
This system is always integrable (see, e.g., \cite{smirnov98}). The TBD criterion, on the other hand, claims that this system is chaotic above $ E = 1/24= 0.04167$ (see \cite{pattanayak97}). \\
Figure \ref{fig:3} is the analogue of figure \ref{fig:1}, at $E = 0.05$, {\it i.e.} above the TBD threshold for the onset of chaos. The figure shows clearly no evidence of chaos according to our criterion. We do not show the counterpart of Fig. \ref{fig:2} for this system, but the results are analog.

\subsection{A potential with mixed convex-concave energy boundary}
\begin{figure}[h]
\includegraphics[width=8.5cm]{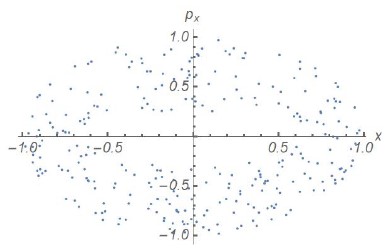}
\caption{Poincarè plot for a trajectory with the potential (\ref{Eq:11}), obtained taking the intersections with the plane $y = 0$ and ${\dot y} <0$. The particle energy is $E = 0.5$ and $a = 2\times 10^{-4}$. }
\label{fig:4}
\end{figure}

As a third example, we choose the potential
\begin{equation}
V = {1 \over 2} (x^2 - y^2) + a \exp(x^2 + y^2)
\label{Eq:11}
\end{equation}
When $a = 0$ the system is integrable, yet it involves exponentially growing solutions (and therefore also relative displacements) along the y-direction. For finite $a $ the potential becomes chaotic: in Fig. \ref{fig:4} we show an example of  Poincar\'e plot for a trajectory, its erratic behaviour being clearly evident. Its chaotic properties are a consequence of its   peculiar geometrical shape: the isopotential contour levels are closed but, unlike most of the potentials commonly encountered, they are not everywhere convex (bulging outwards), rather alternating convex and concave intervals exist.  The role of the energy boundary upon the regular or chaotic kind of motion of a system has often been taken into consideration since it is the region where the strongest acceleration does take place.  In particular, it is critical in connection with
billiards--dynamical systems characterized by step-like potentials: it is well known that the shape of the boundary determines their chaotic or regular nature and that billiards having a concave fraction of the boundary are chaotic \cite{sinai70}. The properties of the billiards are, to some extent, shared by smooth dynamical systems \cite{rapoport07,rod77,li11}. 
 
\begin{figure}[h]
\includegraphics[width=8.5cm]{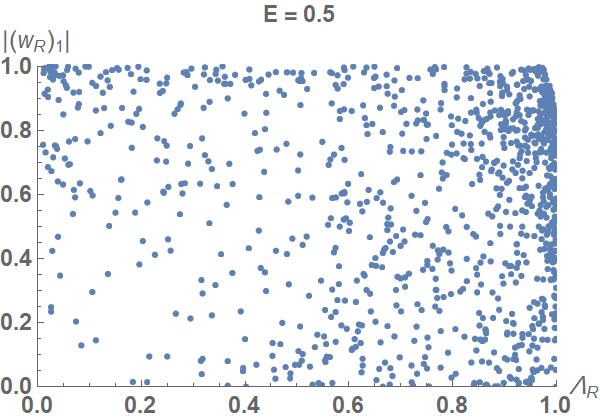}
\caption{The figure is the counterpart of Figs. \ref{fig:1},\ref{fig:3}. The energy is $E = 0.5$, and $a = 2\times 10^{-4}$. The points are scattered throughout the whole available interval, consistently with our guess about chaotic dynamics. }
\label{fig:5}
\end{figure}

We plot in Fig. \ref{fig:5} the equivalent of figs. \ref{fig:1}, \ref{fig:3}. This time, $|({\bf w}_R)_1| $ is wildly fluctuating over the whole $(0,1)$ range, providing a further confirmation to our claims. 

\section{Conclusions}
For Hamiltonian systems of the kind (\ref{eq:1}), with separate kinetic and potential terms, the computation of the eigenvalues in (\ref{Eq:4}) drops the momenta, only the position coordinates are involved. On the one hand, this entails that the stability analysis is carried out over a subspace of dimension one-half of the original space; on the other hand, one does not retain information on the whole dynamics.        
 The shortcoming of the TBD criterion, thus, is that it cannot distinguish between couples of trajectories that are informative about the dynamics of the system, and trajectories that are not, either because do not lie both on the same energy surface, or because are both aligned along the common flow.  \\
In this work we recalled, following the previous paper \cite{kosloff81}, that chaotic systems are those that feature a positive exponential rate of dilatation along the constant-energy hypersurface when the stretching itself is appreciably aligned along directions perpendicular to the local flow. This qualitative observation alone is sufficient to give a justification as to why the TBD criterion blatantly fails when applied to one-dimensional systems. For systems of larger dimensionality, we showed how it is possible to implement these constraints into an analytical algorithm, refinement of the simpler TBD one. In it, besides the mere inventory of the rate of dilatation, also the direction must be taken into account. \\
The resulting algorithm shares several affinities with another index for chaos detection in dynamical systems: the Orthogonal Finite Lyapunov Index (OFLI) \cite{fouchard02}. The only (nontrivial) difference between the two methods lies in the ``frozen coordinates'' hypothesis at the basis of TBD criterion, which allows to decouple the computation of the stretching rate from the particle trajectory.  \\
In this paper, we limited our consideration to two-dimensional systems; There are no conceptual obstacles preventing the same analysis from being applied to higher-dimensional systems, but the computational burden grows accordingly, as well as the difficulty of visualizing and interpreting the results: we leave to further studies the investigation of three-(or higher)-dimensional systems.  \\ 
The amended criterion turns out to correctly catalog the results in all the test cases considered, although--so far--it is a fuzzy rather than an exact criterion. It is obvious, from the inspection of figures \ref{fig:1} and \ref{fig:3}, that regular motion does not demand strictly $|({\bf w}_R)_1| = 1$: a small departure is tolerable. Through visual inspection it seems straightforward to distinguish between regular and chaotic cases. but yielding a quantitative estimate for the maximum allowable amount of this departure is a result that needs further research.  
    
\section*{Acknowledgments}
Barbara Momo read carefully a draft of this work, making several useful suggestions. 
F.S. acknowledges fruitful discussions with D.F. Escande and I. Predebon. 
L.S. is partially supported by the European Quantum Flagship 
Project "PASQuanS 2", by the European Union-NextGenerationEU within 
the National Center for HPC, Big Data and Quantum Computing 
[Project No. CN00000013, CN1 Spoke 10: “Quantum Computing”], 
by the BIRD Project "Ultracold atoms in curved geometries" of the 
University of Padova, and by “Iniziativa Specifica Quantum” of INFN.


\begin{thebibliography}{99}


\bibitem{book0-chaos} M.C. Gutzwiller,
  Chaos in Classical and Quantum Mechanics (Springer, 1990). 

\bibitem{book1-chaos}  J. C. Sprott, Chaos and Time-Series Analysis
  (Oxford Univ. Press, 2001). 

\bibitem{book2-chaos} C. Skokos , G.A. Gottwald, and Jacques Laskar (Eds), 
  Chaos Detection and Predictability, Lecture Notes in Physics
  (Springer, 2016). 


\bibitem{sali} C. Skokos, J. Phys. A {\bf 34}, 10029 (2001)

\bibitem{geodes} C.L.Vozikis, H. Varvoglis, K. Tsiganis, A\&A {\bf 359}, 386 (2000)

\bibitem{fli}  C. Froeschlé, E. Lega and R. Gonzi R., Celest. Mech. Dyn. Astron. {\bf 67}, 41 (1997) 


\bibitem{pd2005} G.A. Gottwald and I. Melbourne,
  Physica D {\bf 212}, 100 (2005)

\bibitem{pd2021} P.M. Cincotta, C.M. Giordano, R.A. Silva,
  and C. Beauge, Physica D {\bf 417}, 132816 (2021)

\bibitem{pd2022} J. Daquin, R. Pedenon-Orlanducci,
  M. Agaoglou, G. Garcia-Sanchez, and A.M. Mancho, 
  Physica D {\bf 442} 133520 (2022)

\bibitem{pd2023} S. Zimper, A. Ngapasare, M. Hillebrand, M. Katsanikas,
S.R. Wiggins, and C. Skokos, Physica D {\bf 453} 133833 (2023)


\bibitem{codaccioni82} J.P. Codaccioni, F. Doveil, D.F. Escande, Phys. Rev. Lett. {\bf 49}, 1879 (1982) 

\bibitem{okubo70} A. Okubo, Deep Sea Res. {\bf 17}, 445 (1970)

\bibitem{toda74} M. Toda, Phys. Lett. {\bf 48A}, 335 (1974)

\bibitem{brumer76} P. Brumer and J. Duff, J. Chem. Phys. {\bf 65}, 3566 (1976) 

\bibitem{enz75} C.P. Enz, M.O. Hongler and C.V. Quach Thi, Helv. Phys. 
Acta {\bf 48}, 787 (1975)

\bibitem{nunez90} H.N. N\'unez-Y\'epez, A.L. Salas-Brito, C.A. Vargas, L. Vicente, Phys. Lett. A {\bf 145}, 101 (1990)

\bibitem{oloumi99} A. Oloumi, D. Teychenn\'e, Phys. Rev. E {\bf 60}, R6279 
(1999)

\bibitem{watabe95} T. Watabe, Phys. Lett. B {\bf 343}, 254 (1995)

\bibitem{kuvnishov02} V.I. Kuvnishov and A.V. Kuzmin, Phs. Lett. A {\bf 296} , 82 (2002)

\bibitem{kuvnishov08} V.I. Kuvnishov and V.A. Piatrou, Phys. Part. Nuclei 
Lett. {\bf 5}, 72 (2008)

\bibitem{li95} Li Junqing, Zhu Jieding, Gu Jinnan, Phys. Rev. B {\bf 52}, 
6458 (1995)


\bibitem{manfredi97} V.R. Manfredi and L. Salasnich, Mod. Phys. Lett. A 
{\bf 12}, 1851 (1997) 

\bibitem{letelier11} P.S. Letelier, J. Ramos-Caro, F. L\'opez-Suspes, Phys. 
Lett. A {\bf 375}, 3655 (2011)

\bibitem{ghosh14} A. Ghosh {\it et al},  Chaos {\bf 24}, 013117 (2014)

\bibitem{chang14} Yu-L. Chang, Lie-Y. Oey, Ocean Dyn. {\bf 64}, 259 (2014)

\bibitem{beron19} F.J. Beron-Vera, A.Hadjighasem, Q. Xia, M.J. Olascoaga, and G. Haller, PNAS {\bf 116}, 18251 (2019)

\bibitem{shivamoggi22} B.K. Shivamoggi, G.J.F. van Heijst  and L.P.J. Kamp, 
Fluid Dynamics Res {\bf 54}, 015505 (2022)

\bibitem{benettin77} G. Benettin, R. Brambilla, L. Galgani, Physica {\bf 87A}, 
381 (1977)

\bibitem{pattanayak97} A.K. Pattanayak and W.C. Schieve, Z. Naturforsch. 
{\bf 52a}, 34 (1997)

\bibitem{smirnov98} R.G. Smirnov, Appl. Math. Lett. {\bf 11}, 71 (1998)

\bibitem{cerjan79} C. Cerjan, W.P. Reinhardt, J. Chem. Phys. {\bf 71}, 1819 
(1979)

\bibitem{kosloff81} R. Kosloff and S.A. Rice, J. Chem. Phys. {\bf 74}, 1947 
(1981)

\bibitem{zotos15} E.E. Zotos, Nonlinear Dyn. {\bf 79},  1665 (2015)


\bibitem{sinai70} Ya.G. Sinai, Russian Math. Surveys {\bf 25},  137 (1970)

\bibitem{rapoport07} A. Rapoport, V. Rom-Kedar, D. Turaev, Commun. Math. Phys. 
{\bf 272}, 567 (2007)

\bibitem{rod77} D.L. Rod, G. Pecelli, R.C. Churchill, Jour. Diff. Eq. 
{\bf 24}, 329 (1977)

\bibitem{li11} J. Li, S. Zhang, Phys. Lett. A {\bf 375}, 1710 (2011)

\bibitem{fouchard02} M.Fourchard, E. Lega, C.Froeschlé, and C.Froeschlé, Celestial Mech. Dyn. Astron. {\bf 83}, 205 (2002)









\end{thebibliography}
\end{document}